\documentclass[preprint2, tighten]{aastex6}
\usepackage{amsmath}
\begin{document}

\title{A first view with GAIA on KIC 8462852 -- \\distance estimates and a comparison to other F stars}
\shorttitle{A first view with GAIA on KIC 8462852}
\shortauthors{Angerhausen \& Hippke}

\author{Michael Hippke}
\email{hippke@ifda.eu}
\affil{Luiter Stra{\ss}e 21b, 47506 Neukirchen-Vluyn, Germany}

\author{Daniel Angerhausen}
\affil{NASA Postdoctoral Program Fellow, NASA Goddard Space Flight Center, Greenbelt, MD 20771, USA}
\email{daniel.angerhausen@nasa.gov}

\begin{abstract}
Distance estimates from Gaia parallax and expected luminosities are compared for KIC 8462852. Gaia DR1 yields a parallax of $2.55\pm0.31$mas, that is a distance of $391.4\substack{+53.6 \\ -42.0}$pc, or $391.4\substack{+122.1 \\ -75.2}$pc including systematic uncertainty. The distance estimate based on the absolute magnitude of an F3V star and measured reddening is $\sim454\pm35$pc. Both estimates agree within $<1\sigma$, which only excludes some of the most extreme theorized scenarios for KIC 8462852. Future Gaia data releases will determine the distance to within 1\% and thus allow for the determination a precise absolute luminosity.
\end{abstract}

\section{Introduction}
The space mission Gaia is currently surveying the entire sky and repeatedly observing the brightest one billion objects, down to 20th magnitude, on a 5-year mission \citep{2016arXiv160904153G}. The telescope collects data providing absolute astrometry, broad-band photometry, and low-resolution spectro-photometry \citep{2012Ap&SS.341...31D}. The first public data was released on 14 September 2016 (DR1, \citet{2016arXiv160904303L}), and contains the five-parameter astrometric solution: positions, parallaxes and proper motions for stars in common between the Tycho-2 Catalogue \citep{2000A&A...355L..27H} and Gaia \citep{2015A&A...574A.115M}.

As an immediate application, we analyze Gaia's distance estimate for KIC 8462852 (TYC 3162-665-1). This object is an F3 main-sequence star, which was observed by the NASA Kepler mission \citep{2010Sci...327..977B} from April 2009 to May 2013. An analysis by \citet{2016MNRAS.457.3988B} shows inexplicable series of day-long brightness dips up to 20\%. The behavior has been theorized to originate from a family of large comets \citep{2016ApJ...819L..34B}, or signs of a Dyson sphere \citep{2016ApJ...816...17W}. Subsequent analysis found no narrow-band radio signals \citep{2016ApJ...825..155H} and no periodic pulsed optical signals \citep{2016ApJ...825L...5S,  2016ApJ...818L..33A}. The infrared flux is equally unremarkable \citep{2015ApJ...815L..27L, 2015ApJ...814L..15M,  2016MNRAS.458L..39T}. Recently, the star has been claimed to dim by $0.16$mag ($\sim14\%$) between 1890 and 1990 \citep{2016ApJ...822L..34S}, and lost $\sim3\%$ of brightness during the 4.25yrs of Kepler mission \citep{2016arXiv160801316M}. The century-long dimming has been challenged by \citet{2016ApJ...825...73H} and \citet{2016arXiv160502760L}. To resolve the controversy whether this star has dimmed by $\sim20\%$ over 130 years (and perhaps more so earlier), a precise distance and therefore a precise absolute luminosity would be very helpful.

\begin{table}
\center
\caption{Gaia data DR1\label{tab:gaia}}
\begin{tabular}{lcccc}
\tableline
Parameter & Value \\
\tableline
Identifier & TYC 3162-665-1 \\
Source ID & 2081900940499099136\\
G mag & 11.685 ($n=140$)\\ 
Parallax & $2.555\pm0.311$mas ($n=109$)\\ 
Distance & $391.4\substack{+53.6  \\ -42.0}$pc only random uncertainty\\
         & $391.4\substack{+122.1 \\ -75.2}$pc incl. systematic uncertainty\\ 
\tableline
\end{tabular}
\end{table}

\begin{figure*}
\includegraphics[width=\linewidth]{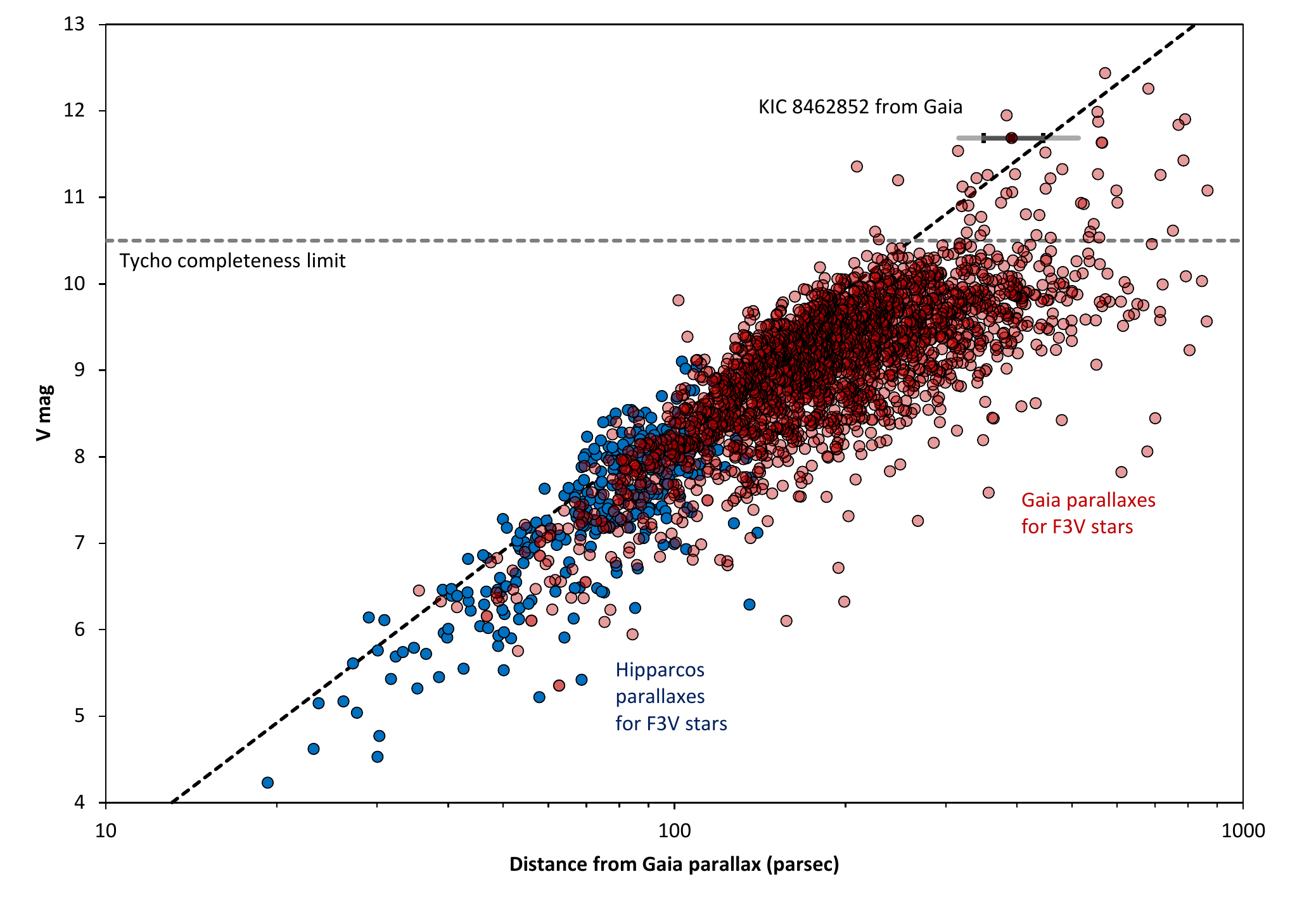}
\caption{\label{fig:parallaxes}Left: Distance-luminosity relation from Hipparcos (blue) and Gaia (red). The dashed line is based on the expected absolute luminosity plus 0.11 mag extinction (Section~\ref{sec:abs}). KIC 8462852 is shown with a black symbol in the upper right corner. Its symbol size represents an uncertainty in brightness of 20\%, corresponding to the deepest dip recorded during the Kepler mission. The distance uncertainty is from Gaia's parallax, with (gray) and without (black) the systematic uncertainty. The non-Gaussian spread of stars is mainly caused by higher reddening values.}
\end{figure*}

\section{Method}
As the absolute luminosity of stars is expected to be known and constant within a few percent, only a reliable distance is required to compare luminosity expectations, and reality. Gaia's final data release will deliver parallaxes for stars at 15th magnitude with an accuracy of $25\mu$as. For a star at a distance of 400 parsec, for example, the parallax would be $1/400=0.0025$ arcsec, or $2500\mu$as. Consequently, Gaia can determine the distance to a star at 15th magnitude at 400 parsec to within 1\% \citep{2012Ap&SS.341...31D}.

\section{Results}
\label{sec:results}

\subsection{Distance estimate from Gaia's parallax}
The distance estimate was downloaded from the \href{http://gea.esac.esa.int/archive/}{Gaia Science Mission Website}. For TYC 3162-665-1 = KIC 8462852, the Data Release 1 (DR1) Catalogue gives $2.55\pm0.31$mas for the distance, i.e. $391.4\substack{+53.6 \\ -42.0}$pc (Table~\ref{tab:gaia}). As expected, the uncertainty in DR1 is higher than predicted for the full mission, since Gaia has so far only conducted part of its full mission. The current DR1 result is based on a total of 140 photometric  and 116 astrometric measurements for this star, of which the pipeline flagged 109 astrometric measurements as ``good''.

Gaia's DR1 potentially contains systematic uncertainties\footnote{See \url{http://www.cosmos.esa.int/web/gaia/dr1}} that are not yet fully characterized but that could reach $300\mu$as \citep{2016arXiv160904303L}. In that case, the distance would be $391.4\substack{+122.1 \\ -75.2}$pc.

\subsection{Distance estimate from absolute magnitudes}
\label{sec:abs}
\citet{2016MNRAS.457.3988B} measure the apparent magnitude of KIC 8462852 as V = 11.705. They take the absolute magnitude of an F3V star as V = 3.08 \citep{2013ApJS..208....9P}, resulting in a (reddened) distance modulus of 8.625. They derive a de-reddened distance of 454pc with $E(B − V ) = 0.11\pm0.03$ (their section 2.4).

The uncertainty in this estimate has to be aggregated from the uncertainty in the apparent brightness ($<1\%$, but dips might produce an offset), the reddening (0.03mag), and the absolute magnitude from stellar models and parallax calibrations ($<0.1$mag, e.g. \citet{1997ApJ...491..749G,1998A&A...333..231B,2009A&A...497..209V}). A conservative total uncertainty of 0.15mag would correspond to a brightness uncertainty of 20\%, and a distance uncertainty of 35pc over 450pc.

\subsection{Comparison stars}
We compare stars which have been identified in the literature as spectral type F3V \citep{2014yCat....1.2023S}. Parallaxes from Hipparcos \citep{1997A&A...323L..49P} and Gaia are used to calculate and plot distance versus apparent magnitude (Figure~\ref{fig:parallaxes}). For the apparent magnitude, Gaia's $G$-passband photometry is broad and covers the range from 330--1000nm \citep{ 2006MNRAS.367..290J,2010A&A...523A..48J} with a peak at $\sim700$nm. The more traditional $V$-magnitude is centered at $\sim550$nm. We have generated the distance versus appararent magnitude figure for both bandpasses, with virtually identical results. Therefore, we show the Figure for the more traditional $V$-band magnitudes.

From Hipparcos, we have 1274 distances of F3V stars, of which 318 have parallax uncertainties $<10\%$, which are shown in with blue symbols the Figure. As Gaia's DR1 is based on Tycho-2, only stars from this catalogue are available. Out of these, there are 2296 F3V stars with astrometric solutions in Gaia's DR1. These are shown with red symbols in the Figure. 

An independent spectroscopic analysis of KIC 8462852 by \citet{2015ApJ...815L..27L} resulted in a classification of F2V instead of F3V. We compared the distance-luminosity relation for F2V and F3V stars based on 2042 parallaxes of stars with these classes from Hipparcos. For a given distance, an F2V star is on average $0.03\pm0.03$ mag brighter than an F3V star. We therefore neglect this classification uncertainty.

\subsection{M-dwarf close companion}
\citet{2016MNRAS.457.3988B} detected an M2V companion star, 4 mag fainter, in a distance of 1.96 arc sec from KIC 8462852. It is unclear if the two stars form a binary or are aligned by coincidence, although the latter is unlikely. Proper motions could shed light on this question, but Gaia's DR1 only contains sources which are in the Tycho-2 catalog, which is not the case for the companion.

\section{Discussion}
\label{sec:discussion}
Recently, \citet{2016arXiv160903505W} discussed ``Families of Plausible Solutions to the Puzzle of Boyajian's Star''. Out of these possibilities, the strongest tension is found for the idea of the star being a ``post-merger returning to normal''. This hypothesis describes some sort of coalescence with another body, such as a brown dwarf or planet, leading to a temporary brightening. The star would then be more luminous than we expected from the reddening, and the claimed dimming would be the return to normal brightness. The hypothesis is described to favour distances of $>500$pc, which is in tension with the Gaia DR1 data by $\sim2\sigma$ ($\sim1\sigma$ including systematic uncertainties). All other hypotheses mentioned by the authors are in tension by $<2\sigma$ with the current parallax. From our judgement, none of these ideas can be excluded at present.

\section{Conclusion}
The distance estimates from absolute magnitude, and parallax measurement agree within $<1\sigma$, but with large uncertainties. As of now, we cannot firmly determine the absolute luminosity of this star. This will be possible with future Gaia data which will constraint the distance to KIC 8462852 within 1\%.

\bibliography{gaia}
\bibliographystyle{aasjournal}
\end{document}